\documentclass[trackchanges]{aastex701}
\usepackage{amsmath}

\begin{document}

\title{Influence of Solar Polar Magnetic Fields  on the Propagation of Coronal Mass Ejection}

\correspondingauthor{Liping Yang, Xueshang Feng}
\email{lpyang@swl.ac.cn, fengx@spaceweather.ac.cn}

\author{Xiao Zhang}
\affiliation{State Key Laboratory of Solar Activity and Space Weather, National Space Science Center, Chinese Academy of Sciences, Beijing 100190, People's Republic of China}
\affiliation{College of Earth and Planetary Sciences, University of Chinese Academy of Sciences, Beijing 100049, People's Republic of China}
\email{ }

\author[0000-0003-4716-2958]{Liping Yang}
\affiliation{State Key Laboratory of Solar Activity and Space Weather, National Space Science Center, Chinese Academy of Sciences, Beijing 100190, People's Republic of China}
\email{ }

\author[0000-0001-8605-2159]{Xueshang Feng}
\affiliation{State Key Laboratory of Solar Activity and Space Weather, National Space Science Center, Chinese Academy of Sciences, Beijing 100190, People's Republic of China}
\affiliation{Shenzhen Key Laboratory of Numerical Prediction for Space Storm, School of Aerospace, Harbin Institute of Technology, Shenzhen 518055, People's Republic of China}
\email{ }

\author[0000-0002-1369-1758]{Hui Tian}
\affiliation{School of Earth and Space Sciences, Peking University, 100871 Beijing, People's Republic of China}
\affiliation{State Key Laboratory of Solar Activity and Space Weather, National Space Science Center, Chinese Academy of Sciences, Beijing 100190, People's Republic of China}
\email{ }

\author{Mengxuan Ma}
\affiliation{State Key Laboratory of Solar Activity and Space Weather, National Space Science Center, Chinese Academy of Sciences, Beijing 100190, People's Republic of China}
\affiliation{College of Earth and Planetary Sciences, University of Chinese Academy of Sciences, Beijing 100049, People's Republic of China}
\email{ }

\author[0000-0002-4935-6679]{Fang Shen}
\affiliation{State Key Laboratory of Solar Activity and Space Weather, National Space Science Center, Chinese Academy of Sciences, Beijing 100190, People's Republic of China}
\affiliation{College of Earth and Planetary Sciences, University of Chinese Academy of Sciences, Beijing 100049, People's Republic of China}
\email{ }

\author[0000-0001-8179-417X]{Jiansen He}
\affiliation{School of Earth and Space Sciences, Peking University, 100871 Beijing, People's Republic of China}
\email{ }

\author[0000-0003-3000-2819]{Man Zhang}
\affiliation{State Key Laboratory of Solar Activity and Space Weather, National Space Science Center, Chinese Academy of Sciences, Beijing 100190, People's Republic of China}
\email{ }

\author{Yufen Zhou}
\affiliation{State Key Laboratory of Solar Activity and Space Weather, National Space Science Center, Chinese Academy of Sciences, Beijing 100190, People's Republic of China}
\email{ }

\author{Ziwei Wang}
\affiliation{Shenzhen Key Laboratory of Numerical Prediction for Space Storm, School of Aerospace, Harbin Institute of Technology, Shenzhen 518055, People's Republic of China}
\email{ }

\author{Xinyi Ma}
\affiliation{State Key Laboratory of Solar Activity and Space Weather, National Space Science Center, Chinese Academy of Sciences, Beijing 100190, People's Republic of China}
\affiliation{College of Earth and Planetary Sciences, University of Chinese Academy of Sciences, Beijing 100049, People's Republic of China}
\email{ }

\author{Wangning Zhang}
\affiliation{State Key Laboratory of Solar Activity and Space Weather, National Space Science Center, Chinese Academy of Sciences, Beijing 100190, People's Republic of China}
\affiliation{School of Geophysics and Information Technology, China University of Geosciences (Beijing), China}
\email{ }

\begin{abstract}

Understanding the propagation of coronal mass ejections (CMEs) through interplanetary space is essential for space weather forecasting. Due to observational limitations, measurements of the photospheric polar magnetic fields remain highly uncertain, and their influence on CME propagation in the heliosphere is still poorly quantified. In this study, we systematically investigate how variations in the photospheric polar magnetic fields affect the Sun-Mars propagation of the 4 December 2021 CME using numerical simulations. The results show that stronger polar fields modify the background solar wind, producing higher plasma density, enhanced magnetic field strength, a flattened heliospheric current sheet, and weakened high-speed streams in the ecliptic plane. These changes markedly slow the CME's radial propagation and inhibit its lateral and radial expansion, leading to notably delayed arrivals at BepiColombo and MAVEN/Tianwen‑1.
Quantitatively, an enhancement of the polar magnetic fields with a peak value of 6 G at the pole decreases the mean propagation and expansion speeds by roughly 200 km s$^{-1}$ and halves the CME volume.
Force analysis reveals that strengthening the polar fields produces only minor changes in the internal force balance of the CME, where the thermal pressure gradient force dominates over the Lorentz force, while it strongly affects the forces acting on the CME surface. At large heliocentric distances, the magnetic pressure of the background solar wind becomes comparable to or even exceeds the aerodynamic drag force, producing a strong confining effect that hinders the CME’s motion.
\end{abstract}

%% but this old "keyword" functionality is maintained in Case authors want


\section{introduction} \label{sec:Introduction}

Coronal mass ejections (CMEs) are among the most energetic solar eruptions, propelling magnetized plasma into the heliosphere at speeds ranging from hundreds to thousands of kilometers per second \citep{gopalswamySOHOLASCOCME2009a,chenCoronalMassEjections2011a,manchesterPhysicalProcessesCME2017b,veronigGenesisImpulsiveEvolution2018,zhangEarthaffectingSolarTransients2021a}.  CMEs can strongly disturb the background solar wind, drive shocks, and compress the interplanetary magnetic field \citep{tsurutaniCoronalMassEjection2003}. Earth-directed  CMEs with a substantial southward magnetic field component are particularly geoeffective, frequently triggering major geomagnetic storms that impact technological systems  \citep{fengGPUacceleratedComputingThreedimensional2013,shenStatisticalComparisonICMEs2017,gopalswamySunSpaceWeather2022}. Therefore, understanding CME initiation and heliospheric evolution is crucial for both fundamental solar physics and space weather forecasting.
 
As a CME propagates from the corona into interplanetary space, its dynamical behavior is strongly influenced by the surrounding plasma and magnetic environment. The speed and density structure of the background solar wind, together with the large-scale magnetic topology, interact collectively to regulate the acceleration or deceleration of CMEs \citep[e.g.,][]{manchesterivEruptionBuoyantlyEmerging2004,shenAccelerationDecelerationCoronal2012}, deflection \citep[e.g.,][]{guiQuantitativeAnalysisCME2011,kayGLOBALTRENDSCME2015,yangGlobalMorphologyDistortion2023}, rotation \citep[e.g.,][]{kumarRotationStealthCME2023,kumarInfluenceSolarWind2024,maInterplanetaryRotation20212024}, expansion \citep[e.g.,][]{scoliniExploringRadialEvolution2021a,yangExpansioninducedThreepartMorphology2025}, among other dynamical effects. 
\citet{yangGlobalMorphologyDistortion2023} employed a passive tracer in their simulation to track a CME's journey from the Sun into interplanetary space. Their work demonstrated that the CME's morphology changes from an ellipsoid to a concave shape due to interactions with the bimodal solar wind.
Since the propagation process of CMEs inherently involves the interplay of multiple forces, previous studies indicate that in the low corona, the CME kinematics are primarily governed by the Lorentz force in the interior of CMEs \citep{beinImpulsiveAccelerationCoronal2011c,carleyCORONALMASSEJECTION2012,shenAccelerationDecelerationCoronal2012,linReviewCurrentSheets2015,
zhouStudyExternalMagnetic2017,shenOriginExtremelyIntense2021,ravishankarKinematicsCoronalMass2020,meiNumericalSimulationLeading2023a,caiMHDModellingNearsun2025}, while at larger heliocentric distances, the aerodynamic drag force exerted by the solar wind becomes the dominant factor \citep{cargillAerodynamicDragForce2004,vrsnakForcesGoverningCoronal2006,temmerINFLUENCEAMBIENTSOLAR2011,sachdevaCMEPROPAGATIONWHERE2015,sachdevaCMEDynamicsUsing2017,abu-shaarCoronalMassEjection2025}. 

During their journey, CMEs not only exhibit changes in propagation behavior but also undergo significant structural evolution, with expansion being a key dynamical feature. This expansion is governed by a combination of internal and external forces. Internally, flux-rope models indicate that expansion is primarily driven by the thermal pressure gradient, while the Lorentz force provides confinement \citep{wangAnalyticalModelProbing2009,mishraModelingThermodynamicEvolution2018}. There are studies showing that a higher initial magnetic pressure of CMEs can enhance expansion in the inner heliosphere \citep{lugazInconsistenciesLocalGlobal2020,verbekeOverexpansionCoronalMass2022,zhangStudyingEvolutionICMEs2025}. Externally, expansion is driven by the declining total pressure of the background solar wind, which causes the CME's internal plasma density to drop more rapidly than its surroundings \citep{demoulinCausesConsequencesMagnetic2009,gulisanoGlobalLocalExpansion2010}. When a CME encounters high-density streams (SIRs), or high-speed streams (HSSs), these enhanced external constraints can suppress expansion or induce significant structural deformation \citep{paganoMHDEvolutionFragment2007,mayankSWASTiCMEPhysicsbasedModel2024}. Overall, CME expansion reflects the interplay between internal force balance and external solar wind conditions, leading to marked deviations from self-similar evolution in interplanetary space \citep{kayModelingInterplanetaryExpansion2021,kaySeriesAdvancesAnalytic2023}.

The polar magnetic fields are a fundamental component of the Sun's global magnetism, playing a key role in governing the solar activity cycle \citep{jiangCAUSEWEAKSOLAR2015,charbonneauDynamoModelsSolar2020,yangLongtermVariationSolar2024,yangVariationsVectorMagnetic2025} and heliospheric conditions \citep{yangSimulationUnusualSolar2011,petrieSolarMagnetismPolar2015}. However, these fields remain poorly constrained observationally due to geometric and instrumental limitations, resulting in sparse measurements with large uncertainties \citep{2020Yangb,2020Yanga,2024Yangc}. These uncertain polar field values propagate into coronal and solar wind models, introducing significant uncertainties into the reconstructed coronal structures and solar wind solutions \citep{tokumaruNondipolarSolarWind2009,luhmannSolarWindSources2009,tokumaruSolarCycleEvolution2010,schonfeldSolarPolarFlux2022,huangModelingSolarWind2023,huangAdjustingPotentialField2024}, and causing a substantial underestimation of the magnetic flux at 1 AU \citep{linkerOpenFluxProblem2017,rileyCanUnobservedConcentration2019,linkerOpenFluxProblem2023,shiRefinementGlobalCoronal2024}. Moreover, the polar field exhibits pronounced solar cycle variability, and its strength serves as a key indicator of the overall state of the background solar wind. Observations reveal clear cycle-to-cycle differences: the polar field during Solar Cycle 23 was weaker than that of Cycle 22 \citep{leeEffectsWeakPolar2009,manoharanThreedimensionalEvolutionSolar2012,bilenkoDeterminationCoronalInterplanetary2018}, while the magnetic field strength, proton density, temperature, and total pressure in Solar Cycle 23 exceeded those in Solar Cycle 24 \citep{jianComparingSolarMinimum2011,manoharanThreedimensionalEvolutionSolar2012,mccomasWEAKESTSOLARWIND2013,gopalswamyAnomalousExpansionCoronal2014,gopalswamyPropertiesGeoeffectivenessMagnetic2015,bilenkoRelationsCoronalMass2020,gopalswamySunSpaceWeather2022}. 

Observations further reveal that CME behavior exhibits solar cycle variability, likely linked to changes in the polar magnetic fields. Comparisons between Solar Cycles 23 and 24 indicate that the significantly lower magnetic and thermal pressures in the heliosphere during Cycle 24 led to stronger CME expansion \citep{gopalswamyAnomalousExpansionCoronal2014}. Magnetic clouds in the weak polar field environment of Cycle 24 also exhibited reduced speeds, weaker magnetic fields, and lower expansion speeds compared to those in Cycle 23 \citep{gopalswamyPropertiesGeoeffectivenessMagnetic2015}. Additionally, the weakening of the photospheric polar field has been associated with a higher CME occurrence rate alongside generally weaker CMEs \citep{petrieEnhancedCoronalMass2015,bilenkoRelationsCoronalMass2020}.

Numerical studies have been conducted to understand how the photospheric magnetic field influences the properties of the solar wind and CMEs. The simulations by \cite{yangSimulationUnusualSolar2011} showed that weak polar fields play an important role in producing the distinct features of the solar corona and the solar wind during the unusual solar minimum between Solar Cycles 23 and 24. \cite{huangSolarWindDriven2024} found that solar wind simulations driven by the Air Force Data Assimilative Photospheric flux Transport-Global Oscillation Network Group (ADAPT-GONG) magnetograms generally show better agreement with OMNI observations than those driven by GONG magnetograms. \cite{doi:10.1126/science.adq0872} developed a near-real-time, data-assimilative coronal MHD model and showed that time-dependent assimilation of evolving photospheric magnetic fields can strongly modify the large-scale coronal magnetic topology. \cite{Shi2025ApJ} found that the magnetic structure of the farside active regions can alter the position of the heliospheric current sheet, thereby affecting the global solar coronal structure. \cite{heinemannQuantifyingUncertaintiesSolar2025} further suggested that improving the operational coverage of concurrent solar magnetic field observations could reduce the uncertainty in the solar wind speed by approximately 100 km s$^{-1}$ near 1 AU. Regarding CMEs, \cite{rileySourcesSizesUncertainty2021} analyzed different realizations of the ADAPT model and found that these realizations lead to interplanetary CME arrival time uncertainties of ±7 hr or more, and \cite{jinAssessingInfluenceInput2022} demonstrated that different solar wind input magnetograms can significantly affect the connectivity and properties of CME-driven shocks in global magnetohydrodynamic (MHD) simulations. 

Despite these advances, a key question remains poorly quantified: how do variations in the solar polar magnetic fields specifically modulate CME evolution through interplanetary space? Addressing this gap is becoming increasingly urgent with new observational capabilities, such as Solar Orbiter's out-of-ecliptic orbits \citep{2020Muller} and China's planned Solar Polar-orbit Observatory, which will provide unprecedented face-on imaging of the polar regions \citep{dengProbingSolarPolar2025a}.

In this work, we investigate the role of solar polar magnetic fields in shaping the interplanetary propagation of the 4 December 2021 CME \citep{chiDynamicEvolutionMultipoint2023,chiInterplanetaryCoronalMass2023,maInterplanetaryRotation20212024,yangExpansioninducedThreepartMorphology2025} using MHD simulations. We systematically quantify how polar field strength modulates the background solar wind, the CME’s kinematics and expansion, and the forces governing its evolution. The paper is structured as follows: Section 2 describes the modeling methodology; Section 3 presents the analysis of the solar wind and CME evolution; and Section 4 summarizes and discusses the findings.

\section{Numerical Methods} \label{sec:Methods}

To simulate the propagation of the CME in the solar corona and interplanetary space, we employ a three-dimensional adaptive mesh refinement (AMR) solar–interplanetary spacetime (SIP) conservation element and solution element (CESE) magnetohydrodynamics (MHD) solar wind model. The theoretical foundation and numerical details of the AMR-SIP-CESE model are described in previous studies \citep{fengTHREEDIMENSIONALSOLARWIND2010,fengValidation3DAMR2012b,yangTimedependentMHDModeling2012,fengDatadrivenModelingSolar2015,fengMagnetohydrodynamicModelingSolar2020,yangNumericalMHDSimulations2021,yangGlobalMorphologyDistortion2023}. The specific simulation setup for the 4 December 2021 CME event is described in \cite{maInterplanetaryRotation20212024} and \cite{yangExpansioninducedThreepartMorphology2025}. In brief, our background solar wind is initialized by Parker’s hydrodynamic isothermal solution, combined with a magnetogram from the Global Oscillation Network Group (GONG) for Carrington Rotation 2251 (CR 2251), which establishes a quasi-steady and realistic background for the propagation of the CME. To account for the solar wind heating and acceleration, the energy source term $Q$ in the MHD equations is formulated as: $Q = Q_1 \exp\left(-\frac{r}{L_{Q_1}}\right) + Q_2 \left(\frac{r}{R_S}-1\right) \exp\left(-\frac{r}{L_{Q_2}}\right)$ . Here, $r$ is the heliocentric distance and $R_S$ is the solar radius. $Q_1$ and $Q_2$ represent the heating intensities, while $L_{Q_1}$ and $L_{Q_2}$ denote the corresponding decay heights \citep{fengTHREEDIMENSIONALSOLARWIND2010,fengValidation3DAMR2012b}. Specifically, $Q_2$ is determined by $Q_2 = Q_0 C_a$, where the heating coefficient $C_a$ is derived from the WSA model. The coefficient $C_a$ is defined as $C_a = C'_a / \max(C'_a)$, where the empirical function $C'_a$ is formulated as:$C'_a = \frac{\left(5.8-1.6e^{[1-(\theta_b/8.5)^3]}\right)^{3.5}}{(1+f_s)^{2/7}}$. Here, $\theta_b$ represents the minimum angular distance of the magnetic footpoint to the nearest coronal hole boundary on the solar surface, and $f_s$ denotes the magnetic expansion factor. It is emphasized that $C_a$ is governed by the magnetic field topology and is independent of the magnetic field strength.
The constants in $Q$ are specified as $Q_1 = 9.33 \times 10^{-9} \text{ J m}^{-3} \text{s}^{-1}$ and $Q_0 = 1.24 \times 10^{-7} \text{ J m}^{-3} \text{s}^{-1}$. For the decay heights, we adopt $L_{Q_1} = 1 R_S$ and $L_{Q_2} = 0.8 R_S$. To eliminate the influence of heating variations on the CME, the energy source term $Q$ was intentionally kept spatially fixed across all cases. In our simulations, no other heating terms are applied, and the specific heat ratio is set to $\gamma = 5/3$.

To investigate the influence of the solar polar magnetic fields on CME propagation in interplanetary space, we first use the original photospheric magnetogram as the input to the MHD model to obtain a steady-state background solar wind, defining Case 1 ($B_\textrm{polar}$). By enhancing the polar magnetic field strength in the input magnetogram, we further construct Case 2 ($B_\textrm{polar}+3G$) and Case 3 ($B_\textrm{polar}+6G$), and obtain the corresponding background solar wind solutions. In the numerical implementation, the polar field enhancements are realized by adding a radial magnetic flux \citep{wangWEAKENINGPOLARMAGNETIC2009a,yangSimulationUnusualSolar2011} to the observed magnetogram, defined as $B_\textrm{rInc} = 3 \sin^7\theta$ (G) for Case 2 and $B_\textrm{rInc} = 6 \sin^7\theta$ (G) for Case 3, respectively. As the input magnetogram only provides the radial magnetic field ($B_r$), the transverse components ($B_\theta, B_\phi$) at the inner boundary of our model are set to zero. The added terms (proportional to $\sin^7\theta$) are antisymmetric with respect to the solar equator, which does not result in a net magnetic flux over the entire spherical surface at the inner boundary (i.e., $\oint \mathbf{B} \cdot d\mathbf{S} = 0$).

With the background solar wind established for each case, we insert a force-free spheromak CME model to simulate the event of 4 December 2021 \citep{kataokaThreedimensionalMHDModeling2009,zhouUsing3DMHD2014,scoliniObservationbasedModellingMagnetised2019,zhangThreedimensionalMHDSimulation2019a,koehnSuccessiveInteractingCoronal2022,palmerioModelingCoronalMass2023,scoliniSpheroidCMEModel2024,yangExpansioninducedThreepartMorphology2025,baratashviliInvestigatingImpactDynamic2025}. The initial parameters of the spheromak are constrained by a Graduated Cylindrical Shell (GCS) fit to white-light coronagraph observations \citep{maInterplanetaryRotation20212024}. The CME is introduced into the background solar wind at time $t = 0$, and we use a passive tracer with values greater than zero ($\rho_c > 0$) to mark the CME \citep{yangGlobalMorphologyDistortion2023,yangExpansioninducedThreepartMorphology2025}. The subsequent evolution, including the CME-solar wind interaction and the associated morphological and dynamical changes of the CME, is governed by the full set of MHD equations.

\section{Results} \label{sec:Results}

In this section, we analyze how the 4 December 2021 CME evolves from the corona into interplanetary space, with particular focus on how different polar field strengths modify the background solar wind and subsequently influence the CME’s propagation, morphology, and dynamics.

\subsection{Influence on the CR2251 Solar Wind Structures}

\begin{figure*}[ht!]
\plotone{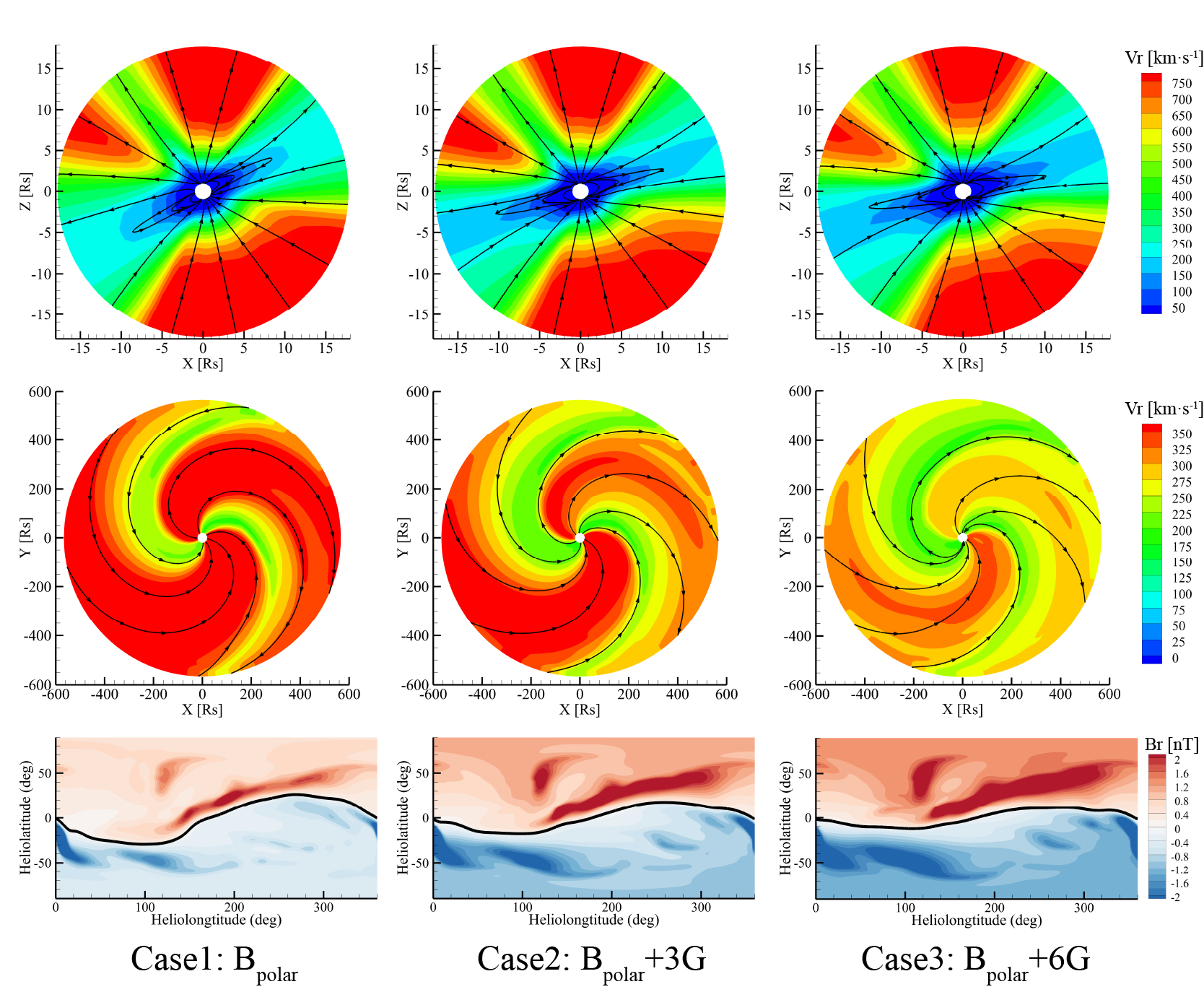}
\caption{Background solar wind structures under the three magnetic configurations: Case 1 ($B_\textrm{polar}$) with the original polar field, Case 2 ($B_\textrm{polar}+3G$) with the enhanced radial polar field of $3 \sin^7\theta$ (G), and Case 3 ($B_\textrm{polar}+6G$) with the enhanced radial polar field of $6 \sin^7\theta$ (G). The top row shows meridional-plane (XZ) slices of the coronal quasi-steady solar wind solution, displaying the radial speed $v_{r}$ and magnetic field lines from 1 to 18 $R_s$. The middle row presents ecliptic-plane (XY) slices of the interplanetary solar wind, displaying the radial speed $v_{r}$ and magnetic field lines from 20 to 560 $R_s$. The bottom row provides synoptic maps of the radial magnetic field $B_r$ at 1~AU, with the black solid line indicating the heliospheric current sheet.
\label{fig:f1}}
\end{figure*}

Figure \ref{fig:f1} shows variations in the background solar wind structure under the three magnetic configurations. As the polar magnetic field strength increases, the magnetic field lines of the streamer deflect towards the equator on the meridional plane (top row). Under the influence of the spatially fixed heating across the three cases, the distribution of radial speed ($v_r$) does not fully follow this trend. In contrast, the high-speed solar wind streams on the ecliptic plane (middle row) weaken. At 1 AU (bottom row), synoptic maps of the radial magnetic field ($B_r$) reveal that a stronger polar field results in a flatter heliospheric current sheet, indicating the growing dominance of the polar magnetic flux in shaping the heliospheric magnetic topology.

\begin{figure*}[ht!]
\plotone{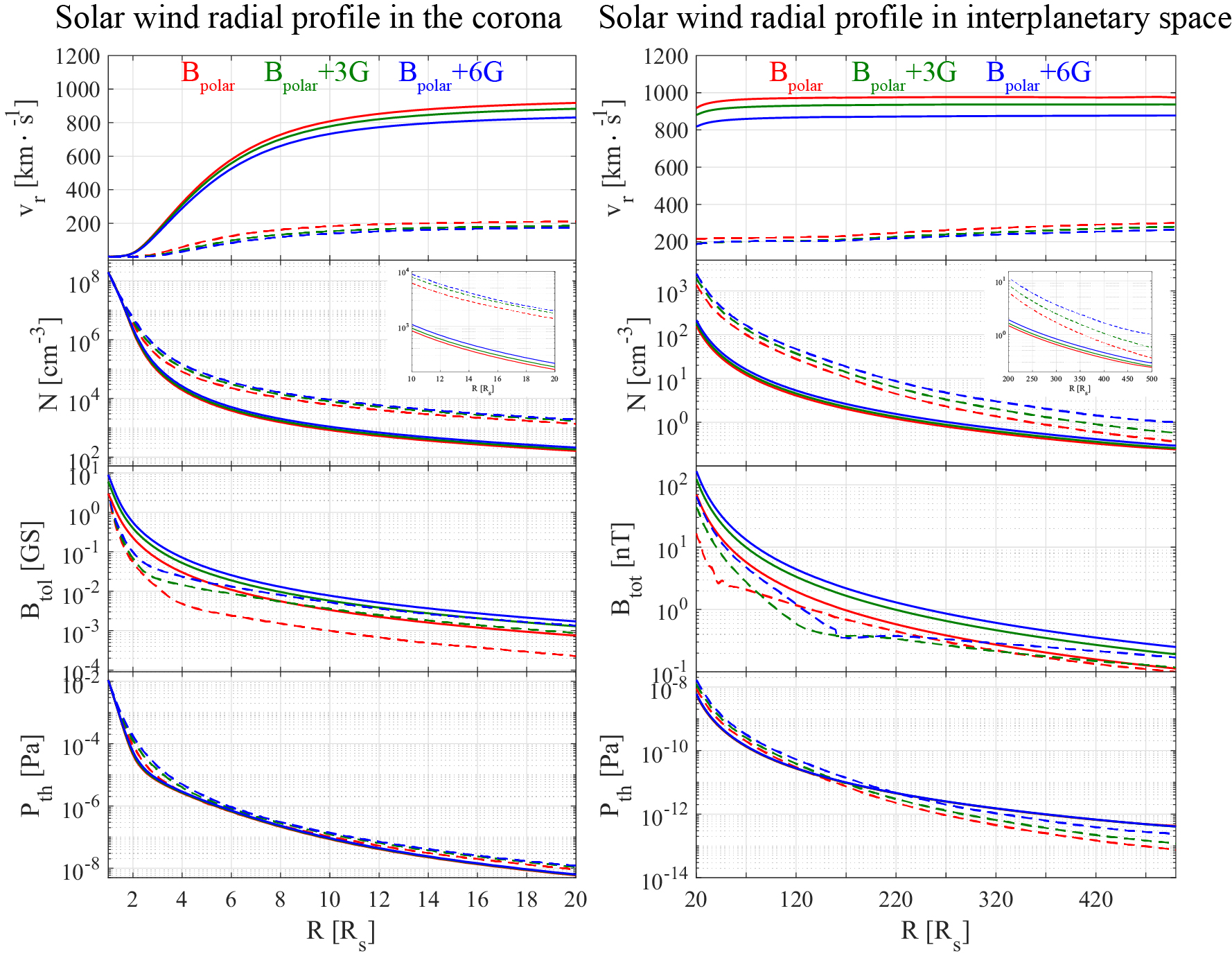}
\caption{Radial profiles of solar wind parameters under the three magnetic configurations: $B_\textrm{polar}$ (red), $B_\textrm{polar}+3G$ (green), and $B_\textrm{polar}+6G$ (blue). High-speed streams are shown by solid lines, and low-speed streams are indicated by dashed lines. The left panels show the radial speed $V_r$, number density $N$, total magnetic field strength $B_\textrm{tot}$, and thermal pressure $P_\textrm{th}$ in the coronal region (from 1 to 20 $R_s$), and the right panels display the corresponding variations in interplanetary space (from 20 to 500 $R_s$). In the second row, the insets in the density $N$ panels present zoomed-in views of the density profiles, highlighting the differences among the three cases.
\label{fig:f2}}
\end{figure*}

Figure \ref{fig:f2} shows the radial evolution of the background solar wind under the three magnetic configurations. To reveal how the polar magnetic fields differentially affect solar wind streams, we separately extract and present the parameters of the high-speed and low-speed streams in Figure \ref{fig:f2}, where the left panels correspond to the coronal region, while the right panels show the evolution in interplanetary space. It can be seen that the strengthening of the polar field leads to a deceleration of the high-speed solar wind, whereas the speed of the low-speed wind is only minimally altered. This change is accompanied by elevated density and magnetic field strength in both wind streams, with thermal pressure showing little variation.   
This is consistent with solar cycle behavior, in which stronger polar fields are associated with higher density and stronger magnetic fields in the solar wind \citep{manoharanThreedimensionalEvolutionSolar2012,mccomasWEAKESTSOLARWIND2013,gopalswamyAnomalousExpansionCoronal2014,gopalswamyPropertiesGeoeffectivenessMagnetic2015}. In our simulations of the three cases, we employ the same solar wind heating and acceleration source, which is empirical and does not explicitly depend on the magnetic field strength \citep{fengTHREEDIMENSIONALSOLARWIND2010,fengValidation3DAMR2012b}. The enhanced polar field leads to higher mass density, and the resulting increase in density naturally reduces the speed of the fast solar wind.

\subsection{Influence on 4 December 2021 CME Propagation}

\begin{figure*}[ht!]
\plotone{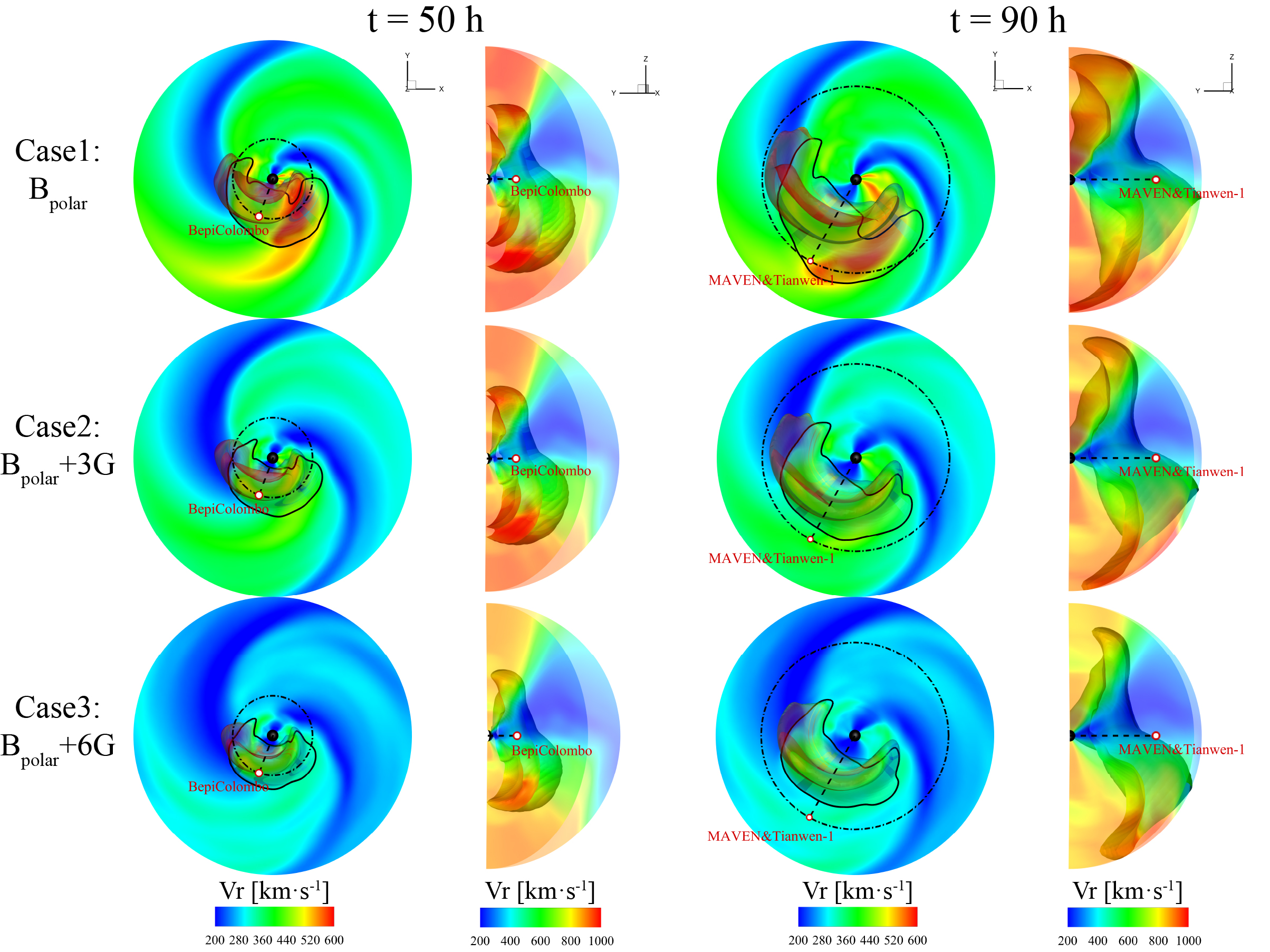}
\caption{Time evolution of CME propagation under the three magnetic configurations: Case 1 ($B_\textrm{polar}$), Case 2 ($B_\textrm{polar}+3G$), and Case 3 ($B_\textrm{polar}+6G$). The left and right parts show the CME structure at $t=50\ \textrm{hr}$ and $t=90 \ \textrm{hr}$, respectively. In each part, the first column illustrates the CME evolution in the ecliptic plane (XY) (20-500 $R_s$), and the second column presents the corresponding meridional-plane slices (20-480 $R_s$). The CME body is outlined by shaded isosurfaces (3D views) and solid-line contours (2D slices), superimposed on the radial speed $V_r$. The dash–dotted lines denote the trajectories of BepiColombo and MAVEN/Tianwen-1, with their positions shown by the red circles.
\label{fig:f3}}
\end{figure*}

Figure \ref{fig:f3} presents the time evolution of CME propagation under the three magnetic configurations, showing that the enhancement of the polar magnetic field modifies both the arrival time and the encounter position of the CME at the spacecraft. At $t=50\ \textrm{hr}$, the relative positions between the CME and BepiColombo already exhibit slight differences among the three cases. By $t=90\ \textrm{hr}$, these differences become more pronounced: in Case 1, the CME has already passed MAVEN/Tianwen-1; in Case 2, it is arriving at MAVEN/Tianwen-1; whereas in Case 3, it has not yet reached MAVEN/Tianwen-1.  

\begin{figure*}[ht!]
\plotone{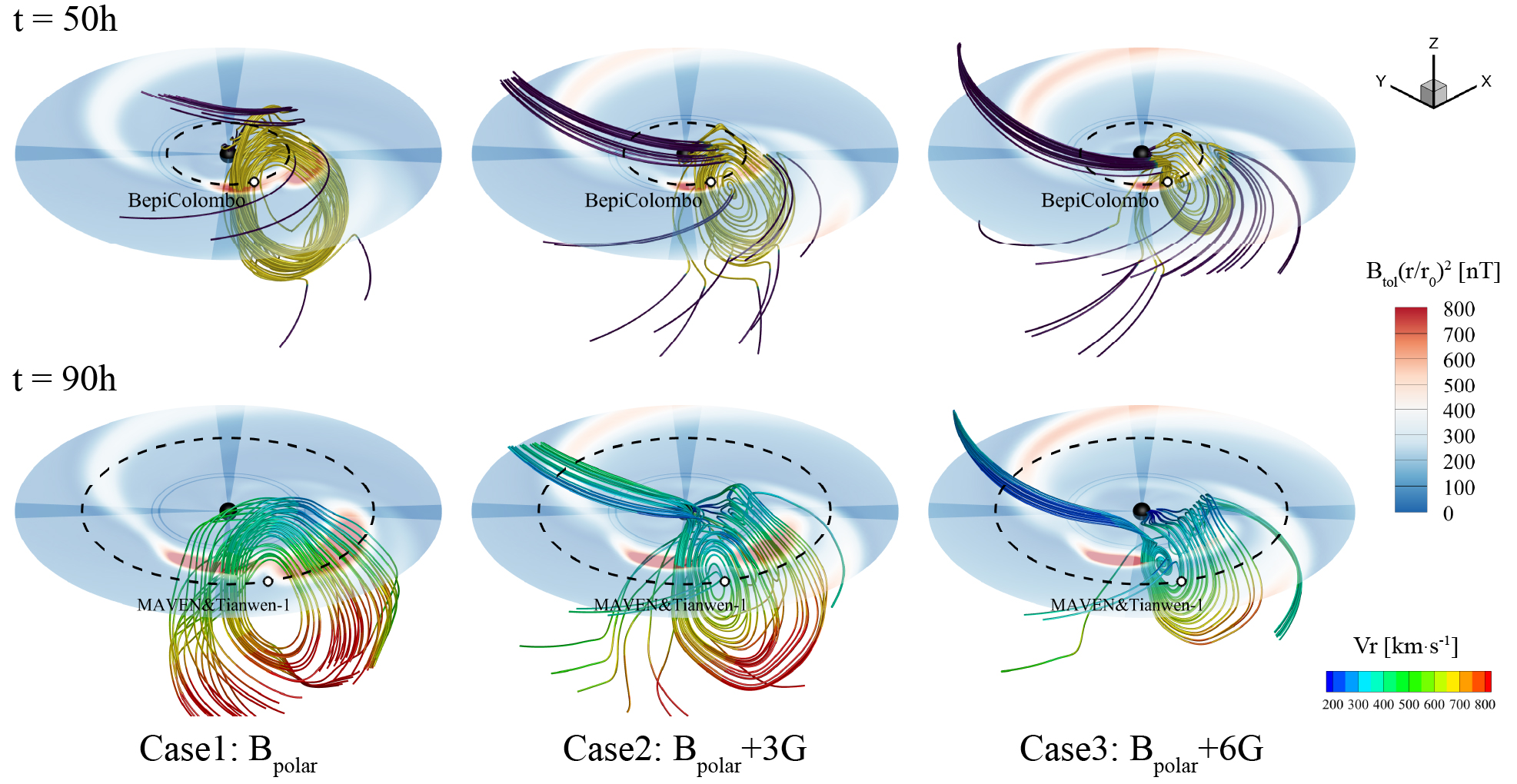}
\caption{Magnetic structure of the simulated CME under the three magnetic configurations: Case 1 ($B_\textrm{polar}$), Case 2 ($B_\textrm{polar}+3G$), and Case 3 ($B_\textrm{polar}+6G$). The top and bottom panels show the CME structure at $t=50\ \textrm{hr}$ and $t=90\ \textrm{hr}$, respectively. A semi-transparent slice on the equatorial plane from 20 to 500 $R_s$ displays the background magnetic field distribution, with the colors indicating the normalized magnetic field strength $B_\textrm{tot}(r/r_0)^2$. In the top panels, the magnetic field lines are colored by $\rho_c$ (yellow for the CME and purple for the background solar wind), while in the bottom panels, they are colored according to the radial velocity $V_r$. The dashed lines denote the trajectories of BepiColombo and MAVEN/Tianwen-1, with their positions shown by the black circles.
\label{fig:f4}}
\end{figure*}

Figure \ref{fig:f4} further illustrates the three-dimensional magnetic structure of the simulated CME under the three magnetic configurations. 
At $t=50\ \textrm{hr}$ (top panels), the CME magnetic structure remains relatively compact, yet the field-line geometry already differs among the cases. A stronger polar magnetic field results in a more confined core.
By $t=90\ \textrm{hr}$ (bottom panels), the CME magnetic structure has expanded considerably, and the distinctions among the cases become more pronounced. In Case 1, the magnetic structure extends farther outward with a broad spatial distribution, whereas in Case 3 it stays compact and closer to the Sun. Notably, in Cases 2 and 3, distinct magnetic tails are observed extending to the upper left. Based on the distributions of $\rho_c$ and $V_r$, these magnetic tails are associated with the background low-speed solar wind stream. As these field lines connect the CME to the background wind, their formation is attributed to magnetic reconnection between the CME magnetic field and the background low-speed solar wind stream. The dashed trajectories of BepiColombo and MAVEN/Tianwen-1 provide spatial reference, highlighting how the polar field strength affects both the spatial extent of the CME’s magnetic structure and its arrival at these spacecraft.

\begin{figure*}[ht!]
\plotone{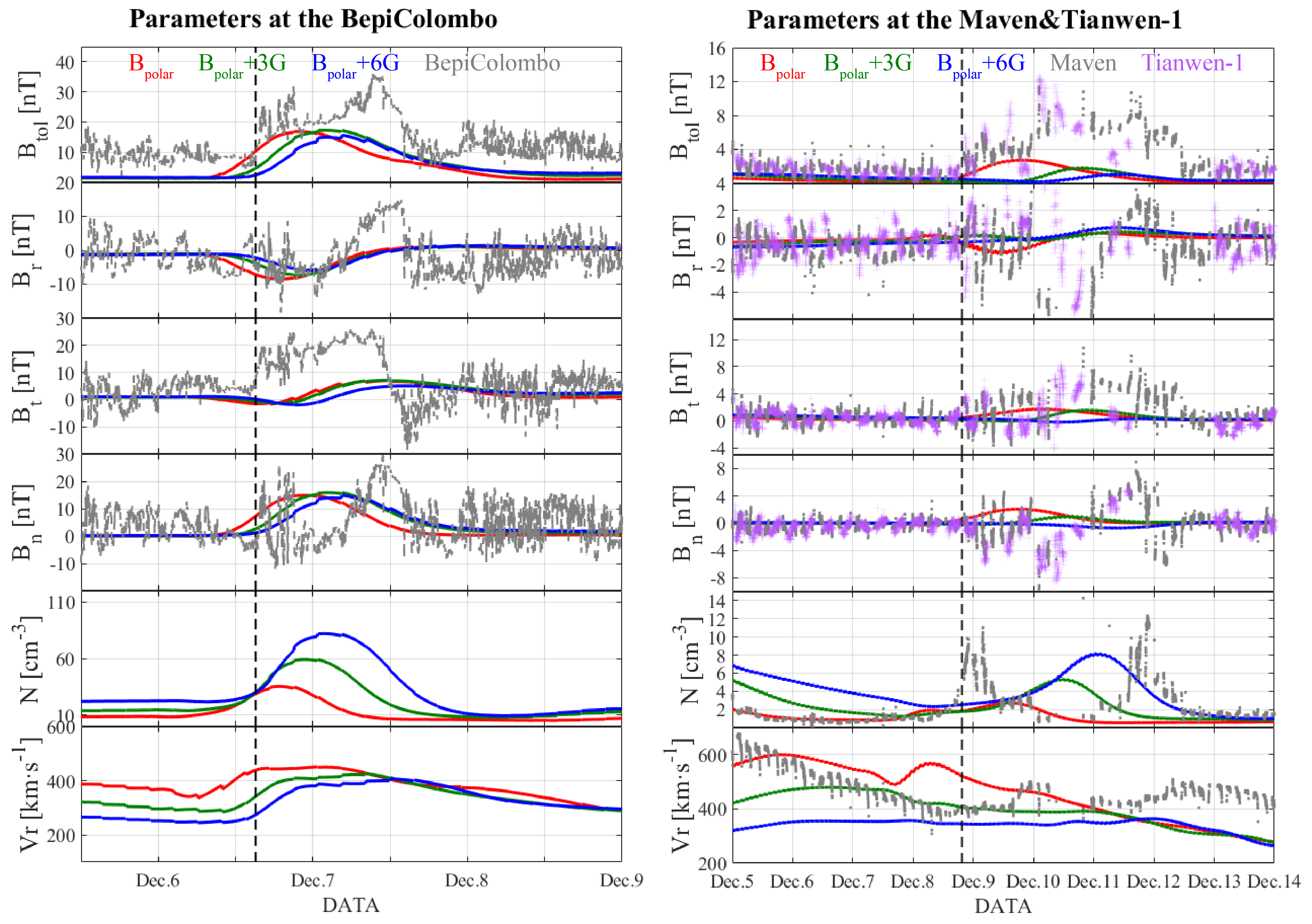}
\caption{Comparison of the simulated and in situ CME signatures at BepiColombo (left panels) and at MAVEN/Tianwen-1 (right panels) under the three magnetic configurations: Case 1 ($B_\textrm{polar}$), Case 2 ($B_\textrm{polar}+3G$), and Case 3 ($B_\textrm{polar}+6G$). For each spacecraft, the panels display time series of the magnetic field magnitude $B_\textrm{tot}$, the magnetic field vector $(B_r, B_t, B_n)$ in RTN coordinates, the number density $N$, and the radial speed $V_r$. The simulated profiles (colored lines: red for Case 1, green for Case 2, and blue for Case 3) are overlaid with the corresponding in situ measurements (gray for BepiColombo; gray and purple for MAVEN and Tianwen-1, respectively). The vertical dashed lines mark the observed CME arrival times at each spacecraft.
\label{fig:f5}}
\end{figure*}

Figure \ref{fig:f5} compares the simulated CME signatures with in situ measurements from BepiColombo and MAVEN/Tianwen-1 under the three magnetic configurations. It is worth noting that both the in situ measurements and the simulation show no distinct shock structure. The absence of a clear shock here is likely attributed to the low speed of the CME near the ecliptic plane, which can be seen in Figure \ref{fig:f3}. At both spacecraft, the arrival time of the simulated CME in Case 1 (red) agrees most closely with observations. In contrast, the simulated CMEs in Case 2 (green) and Case 3 (blue) show progressively greater delays, a trend consistent with their decreasing radial speeds from Case 1 to Case 3. This confirms that a stronger polar field slows the CME. Regarding magnetic field strength, the three cases show similar profiles at BepiColombo (0.67 AU), whereas at the heliocentric distance of MAVEN/Tianwen-1 (1.57 AU), Case 1 maintains a slightly higher field strength than Cases 2 and 3. In contrast, the density profiles reveal a consistent trend at both locations: Case 3 produces the largest enhancement, followed by Case 2 and then Case 1. This demonstrates that a stronger polar field systematically leads to a denser and more compressed sheath during CME propagation.

\subsection{Influence on 4 December 2021 CME Expansion}

\begin{figure*}[ht!]
\plotone{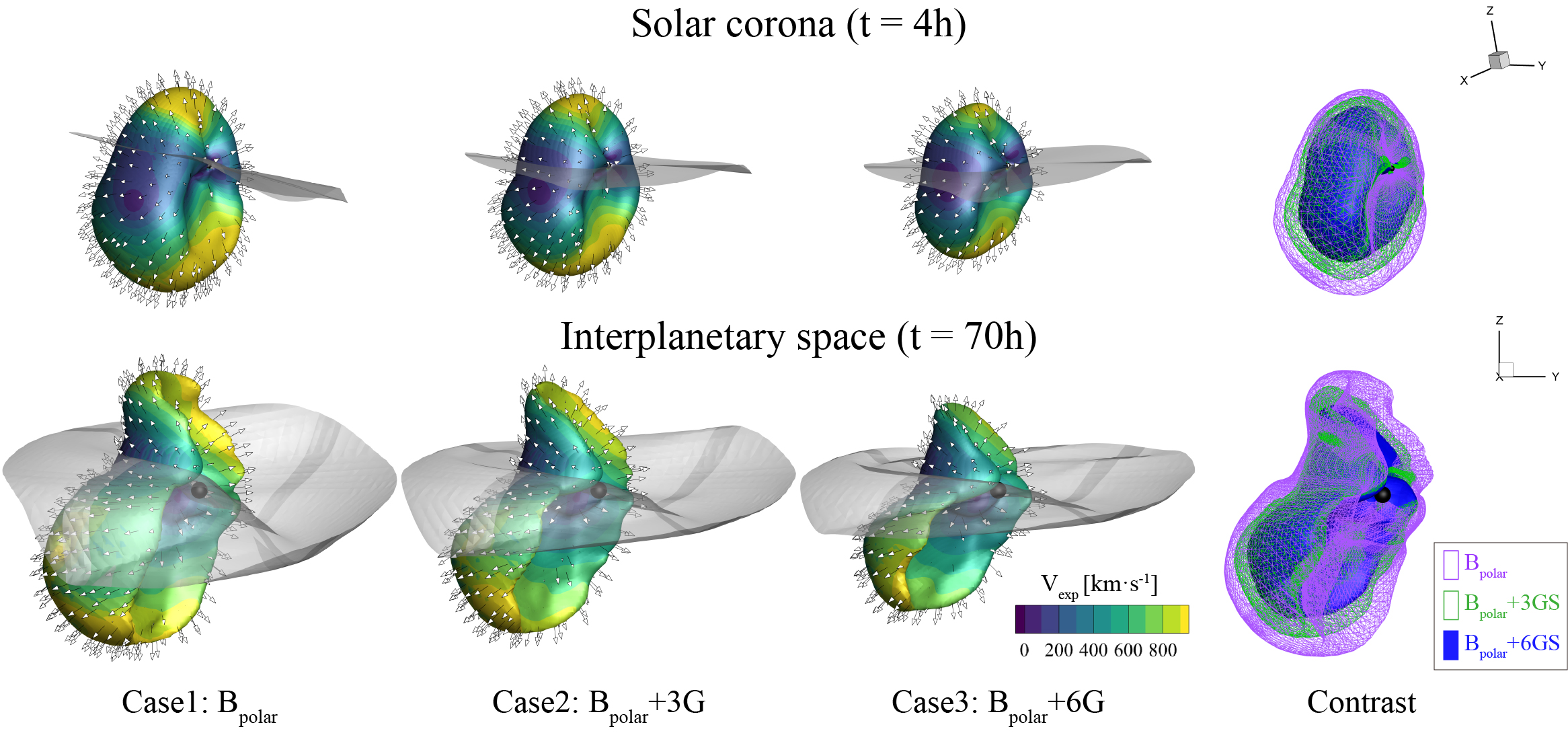}
\caption{Three-dimensional morphology of the CME in the solar corona ($t = 4 \ \textrm{hr}$) and in interplanetary space ($t = 70 \ \textrm{hr}$) under the three magnetic configurations: Case 1 ($B_\textrm{polar}$), Case 2 ($B_\textrm{polar}+3G$), and Case 3 ($B_\textrm{polar}+6G$). The CME structure is marked using the passive tracer $\rho_c$, and represented by the isosurfaces $\rho_c = 10^{-3}$ (solar corona) and $\rho_c = 10^{-5}$ (interplanetary space). The isosurface color represents the expansion speed $V_\textrm{exp}$, and arrows show their directions. The gray semi-transparent surface indicates the heliospheric current sheet, which spans 1–50 $R_s$ in the corona and 20–500 $R_s$ in interplanetary space. The fourth panels provide a composite view that nests the isosurfaces of $\rho_c$ from all three cases, where the purple mesh corresponds to Case 1, the green mesh corresponds to Case 2, and the blue solid surface corresponds to Case 3.
\label{fig:f6}}
\end{figure*}

Figure \ref{fig:f6} illustrates how the CME morphology evolves as it propagates from the low corona to interplanetary space under the three magnetic configurations. The coronal and interplanetary structures are represented by the isosurfaces $\rho_c = 10^{-3}$ (top panels) and $\rho_c = 10^{-5}$ (bottom panels), respectively. 
In both regions, the CME becomes progressively smaller as the strength of the polar magnetic fields increases. This trend is highlighted in the composite view (the fourth panels), where the nested isosurfaces (Case 1 in purple, Case 2 in green, and Case 3 in blue) clearly reveal the differences in size and shape of the CME among the three magnetic configurations. The colors on the isosurfaces represent the expansion speed $V_\textrm{exp}$, which is computed as the residual velocity obtained by subtracting the bulk propagation speed of the CME from the local plasma velocity     \citep{yangExpansioninducedThreepartMorphology2025}.
The expansion speed $V_\textrm{exp}$ displays a systematic weakening from Case 1 to Case 3 with increasing polar field strength. In addition, the gray semi-transparent surface indicates the heliospheric current sheet, which becomes flatter as the polar magnetic field increases.

\begin{figure*}[ht!]
\plotone{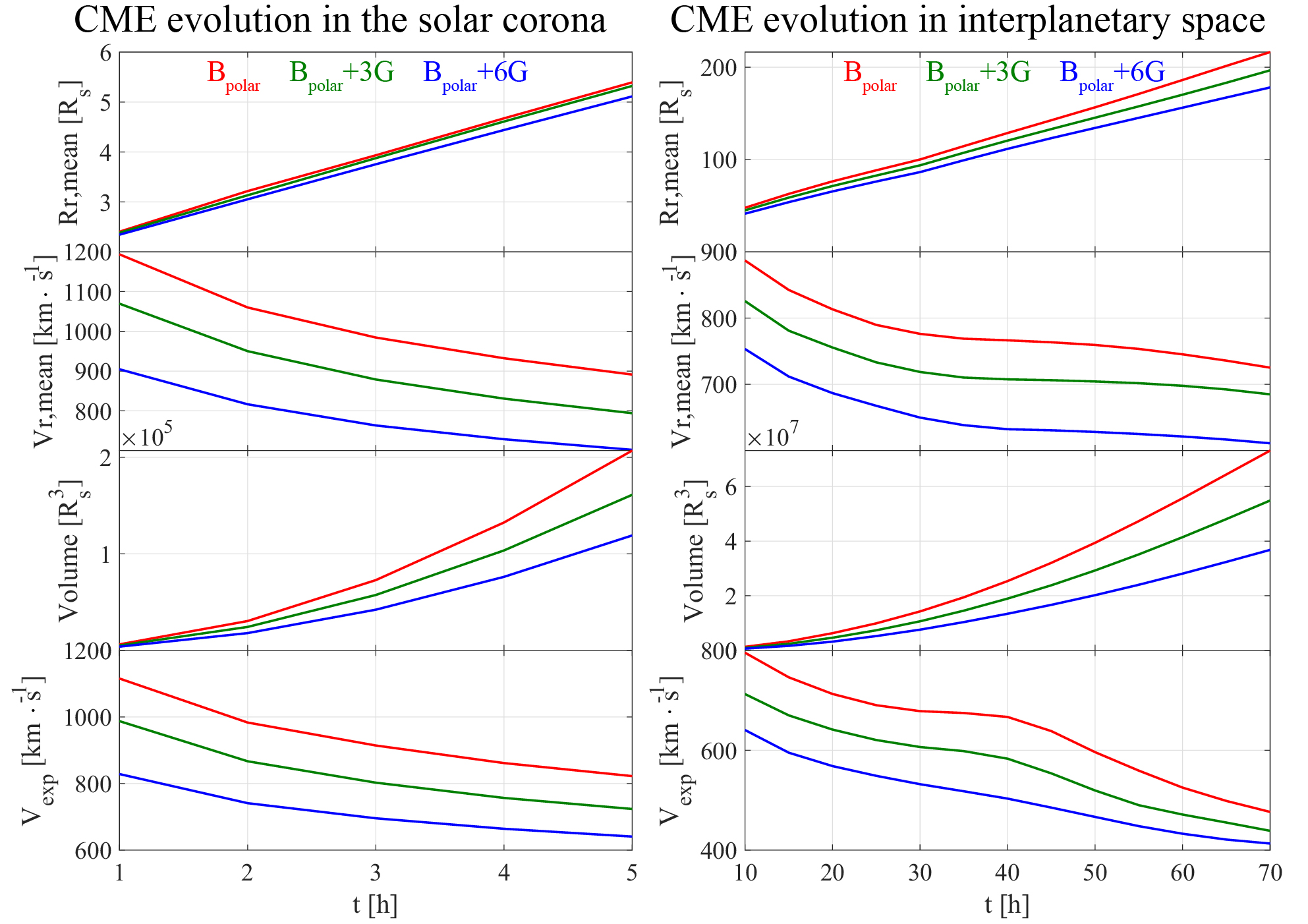}
\caption{Time evolution of the CME in the solar corona (left panels) and in interplanetary space (right panels) under the three magnetic configurations: Case 1 ($B_\textrm{polar}$, red), Case 2 ($B_\textrm{polar}+3G$, green), and Case 3 ($B_\textrm{polar}+6G$, blue). The isosurface represents the CME structure defined by a constant CME tracer $\rho_c$. For each region, the panels show the evolution of four CME parameters: the mean radial distance $R_\textrm{r,mean}$, mean radial speed $V_\textrm{r,mean}$, volume, and expansion speed $V_\textrm{exp}$.
\label{fig:f7}}
\end{figure*}

Figure \ref{fig:f7} shows the temporal evolution of the CME in the solar corona (left panels) and in interplanetary space (right panels) under the three magnetic configurations. The four parameters displayed in the panels—the mean radial distance $R_\textrm{r,mean}$, the mean radial speed $V_\textrm{r,mean}$, volume, and the expansion speed $V_\textrm{exp}$—are all obtained from the grid cells with $\rho_c>0$, which mark the body of the CME. The definitions and computation methods for $R_\textrm{r,mean}$, $V_\textrm{r,mean}$, volume, and $V_\textrm{exp}$ follow the methods described in \citet{yangGlobalMorphologyDistortion2023, yangExpansioninducedThreepartMorphology2025} and are not repeated here. The coronal and interplanetary results reveal a consistent ordering among the three cases: Case 1 reaches the largest radial distance, shows the highest radial propagation speed, attains the greatest volume, and exhibits the highest expansion speed. By contrast, strengthening the polar magnetic field in Case 2 and Case 3 leads to a progressively smaller CME with reduced radial propagation speed, volume, and expansion speed. 
Comparison of Case 1 and Case 3 reveals that a 6 G enhancement of the polar magnetic field reduces the CME’s mean radial propagation and expansion speeds by roughly 200 km s$^{-1}$, and decreases the CME volume to about half of its value without polar field enhancement.

\begin{figure*}[ht!]
\plotone{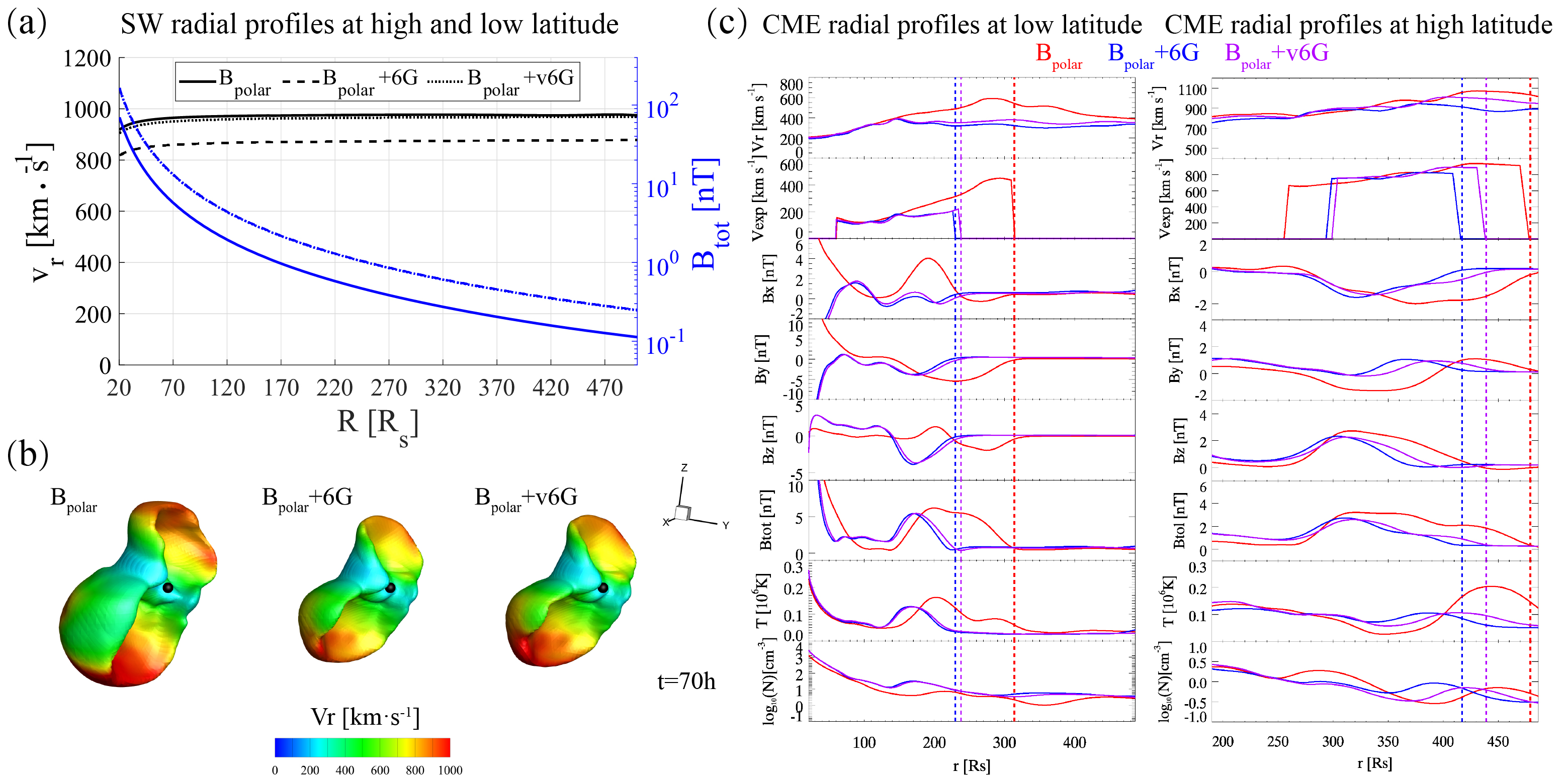}
\caption{Velocity-controlled simulation of CME expansion and propagation under the three magnetic configurations: Case 1 ($B_\textrm{polar}$), Case 3 ($B_\textrm{polar}+6G$), and Case 4 ($B_\textrm{polar}+v6G$, with enhanced polar fields but restored wind speed). (a) Background solar wind radial profiles of $V_r$ (blue) and $B_\textrm{tot}$ (black), where solid, dashed, and dash–dotted lines correspond to Case 1, Case 3, and Case 4, respectively. (b) Three-dimensional CME morphology at $t = 70 \ \textrm{hr}$, colored by radial speed $V_r$. The isosurface represents the CME structure defined by a constant CME tracer $\rho_c$.  (c) Radial profiles of the CME at low latitude (left column) and high latitude (right column). From top to bottom, the panels show $V_r$, $V_\textrm{exp}$, $B_x$, $B_y$, $B_z$, $B_\textrm{tot}$, $T$, and $\log_{10}(N)$. Vertical dashed lines indicate the positions of the CME leading edges.
\label{fig:f8}}
\end{figure*}

To determine whether the reduced CME propagation and expansion in enhanced polar field cases are due to the reduced background solar wind speed (Figures \ref{fig:f1} and \ref{fig:f2}), we conduct an additional test (Case 4, with enhanced polar fields but restored wind speed, denoted as $B_\textrm{polar}+v6G$). Here, we adopt the polar field from Case 3 but adjust the solar wind heating and acceleration parameters to restore the background solar wind speed to a level comparable to Case 1. Specifically, the heating intensity constant $Q_0$ in the energy source term $Q$ was increased from $1.24 \times 10^{-7} \text{ J m}^{-3} \text{s}^{-1}$ to $1.62 \times 10^{-7} \text{ J m}^{-3} \text{s}^{-1}$. As shown in Figure \ref{fig:f8}(a), the radial speed profiles in the high-speed wind of Cases 1 and 4 are closely matched. However, the total magnetic field strength ($B_\textrm{tot}$) in Case 4 remains elevated—similar to Case 3 and higher than in Case 1.

After controlling for solar wind speed, Figure \ref{fig:f8}(b) shows that the CME in Case 4 has a volume comparable to that in Case 3 but remains significantly smaller than in Case 1. Similarly, its radial speed $V_r$ matches that of Case 3 and stays notably lower than in Case 1. These results indicate that the reduced CME expansion and propagation in Case 3 persist when the background solar wind speed is restored to the Case 1 level, demonstrating that the enhanced polar magnetic fields themselves are the primary factor limiting CME motion. Figure \ref{fig:f8}(c) further supports this conclusion through radial profiles at different latitudes. Despite nearly identical solar wind speeds, the CME in Case 4 exhibits slower propagation, weaker expansion, and a clear lag in the leading-edge position relative to Case 1. Together, Figures \ref{fig:f8}(a), \ref{fig:f8}(b), and \ref{fig:f8}(c) demonstrate that the enhanced polar magnetic fields themselves, rather than changes in the background wind speed, are the primary factor limiting the CME expansion and propagation.

\subsection{Influence on 4 December 2021 CME Dynamics}

To investigate the physical mechanism responsible for the slower propagation and weaker expansion of the CME under enhanced polar fields, we examine the forces acting on both the interior and the surface of the CME.

\begin{figure*}[ht!]
\plotone{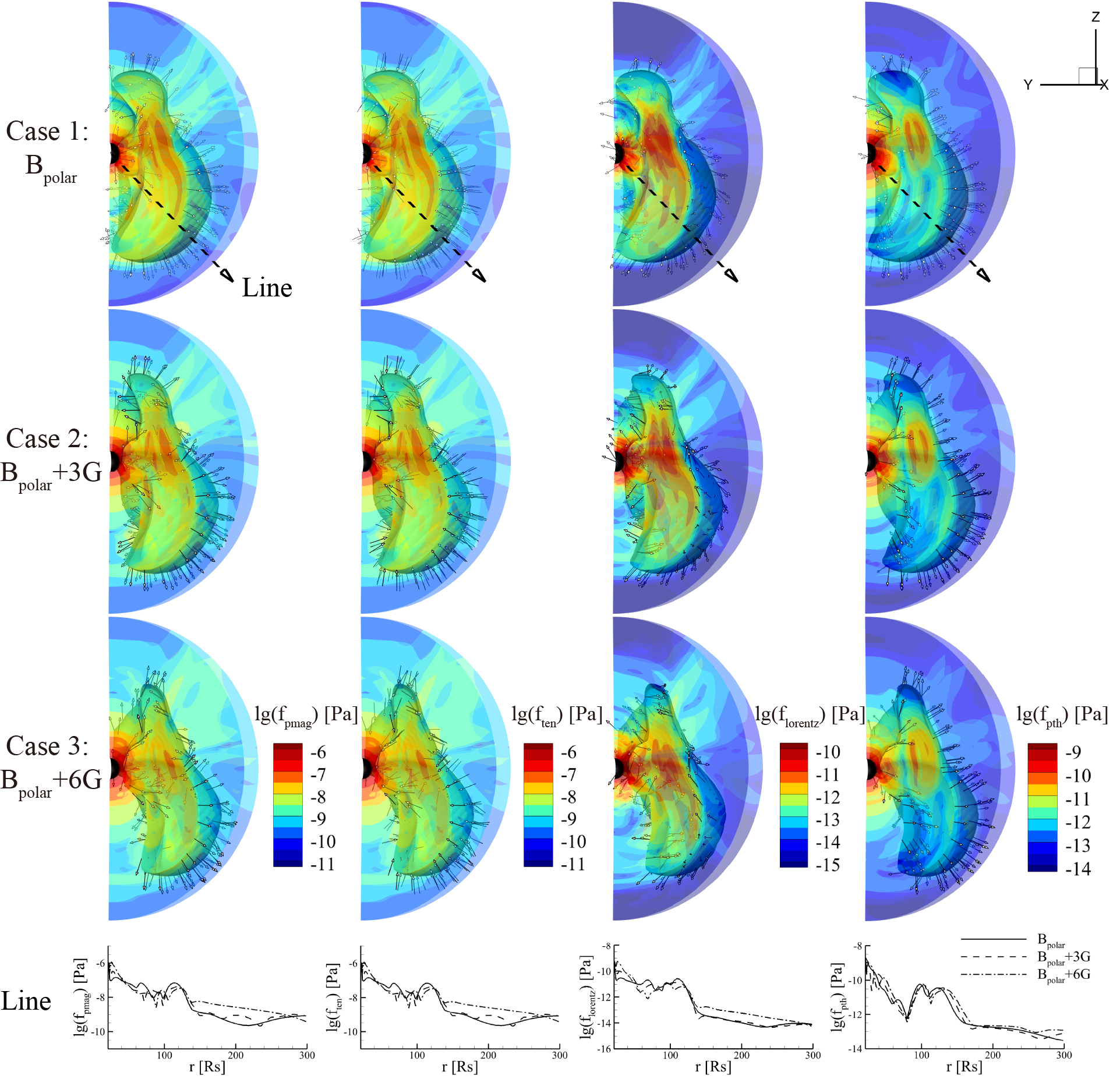}
\caption{Distributions of the internal force densities within the CME under the three magnetic configurations: Case 1 ($B_\textrm{polar}$), Case 2 ($B_\textrm{polar}+3G$), and Case 3 ($B_\textrm{polar}+6G$). The four columns show, from left to right, the magnetic pressure gradient force $\log_{10}(f_\textrm{pmag})$, the magnetic tension force $\log_{10}(f_\textrm{ten})$, the Lorentz force $\log_{10}(f_\textrm{lorentz})$, and the thermal pressure gradient force $\log_{10}(f_\textrm{pth})$. In each panel with a radial range of 20-300 $R_s$, the isosurface represents the CME structure defined by a constant CME tracer $\rho_c$, and colors and black arrows indicate the magnitudes and directions of the forces, respectively. The bottom row shows the corresponding radial profiles of the force densities extracted along the dashed line marked in the upper panels, allowing a quantitative comparison among the three cases.
\label{fig:f9}}
\end{figure*}

Figure \ref{fig:f9} presents the distributions of the magnetic pressure gradient force density $\log_{10}(f_\textrm{pmag})$, the magnetic tension force density $\log_{10}(f_\textrm{ten})$, the Lorentz force density $\log_{10}(f_\textrm{lorentz})$, and the thermal pressure gradient force density $\log_{10}(f_\textrm{pth})$ under the three magnetic configurations. 
Because the CME’s speed, density, magnetic field, and other physical quantities vary with heliocentric distance, here we compare the cases when the CME is at similar heliocentric distances rather than at the same time. The Lorentz force density acting on the CME is defined as $\mathbf{f}_{\textrm{lorentz}} = \mathbf{j} \times \mathbf{B}$, where $\mathbf{B}$ denotes the magnetic field and $\mathbf{j} = \nabla \times \mathbf{B}$ is the current density. The Lorentz force density inside the CME is decomposed into the magnetic tension force density $\mathbf{f}_{\textrm{ten}}=(\mathbf{B}\cdot\nabla)\mathbf{B}
$ and the magnetic pressure gradient force density $\mathbf{f}_{\textrm{pmag}}=-\nabla{{B_{\textrm{tol}}}^2/2}
$ ($B_{\textrm{tol}}$ is the magnitude of $\mathbf{B}$), which together characterize the internal magnetic response of the CME. The detailed definitions and computation procedures for the Lorentz force density, the magnetic tension density, and the magnetic pressure gradient density follow the methodology of \citet{yangExpansioninducedThreepartMorphology2025}. The thermal pressure gradient force density is defined as $\mathbf{f}_{\textrm{pth}}=-\nabla{P_\textrm{th}}$, with $P_\textrm{th}$ being the thermal pressure.

As shown in Figure \ref{fig:f9}, the magnetic pressure gradient force density $f_\textrm{pmag}$ (the first column) and the magnetic tension force density $f_\textrm{ten}$ (the second column) have comparable magnitudes but opposite directions. As shown in Figure \ref{fig:f9}, the magnetic pressure gradient force density $f_{pmag}$ (the first column) and the magnetic tension force density $f_{ten}$ (the second column) have comparable magnitudes but opposite directions. As a result, the net Lorentz force density is non-zero but approximately 4--5 orders of magnitude smaller than either of its individual components. The magnetic pressure gradient force density acts outward and therefore promotes the CME motion, whereas the magnetic tension force density acts inward and suppresses the CME motion. The thermal pressure gradient force (the fourth column) exceeds the Lorentz force (the third column) and thus becomes the dominant internal driver of the CME for the three cases, a behavior consistent with earlier studies \citep{wangAnalyticalModelProbing2009,mishraModelingThermodynamicEvolution2018}. This conclusion is further supported by the radial profiles shown in the bottom row, which are extracted along the dashed line marked in the upper row. These profiles demonstrate that the relative magnitudes and radial trends of the internal force densities remain nearly unchanged among the three cases. Across all three magnetic configurations, increasing $B_\textrm{polar}$ results in only subtle changes in the internal force density distributions, confirming that the reduced propagation and suppressed expansion in the enhanced polar field cases are not primarily caused by modifications of the internal forces within the CME.

\begin{figure*}[ht!]
\plotone{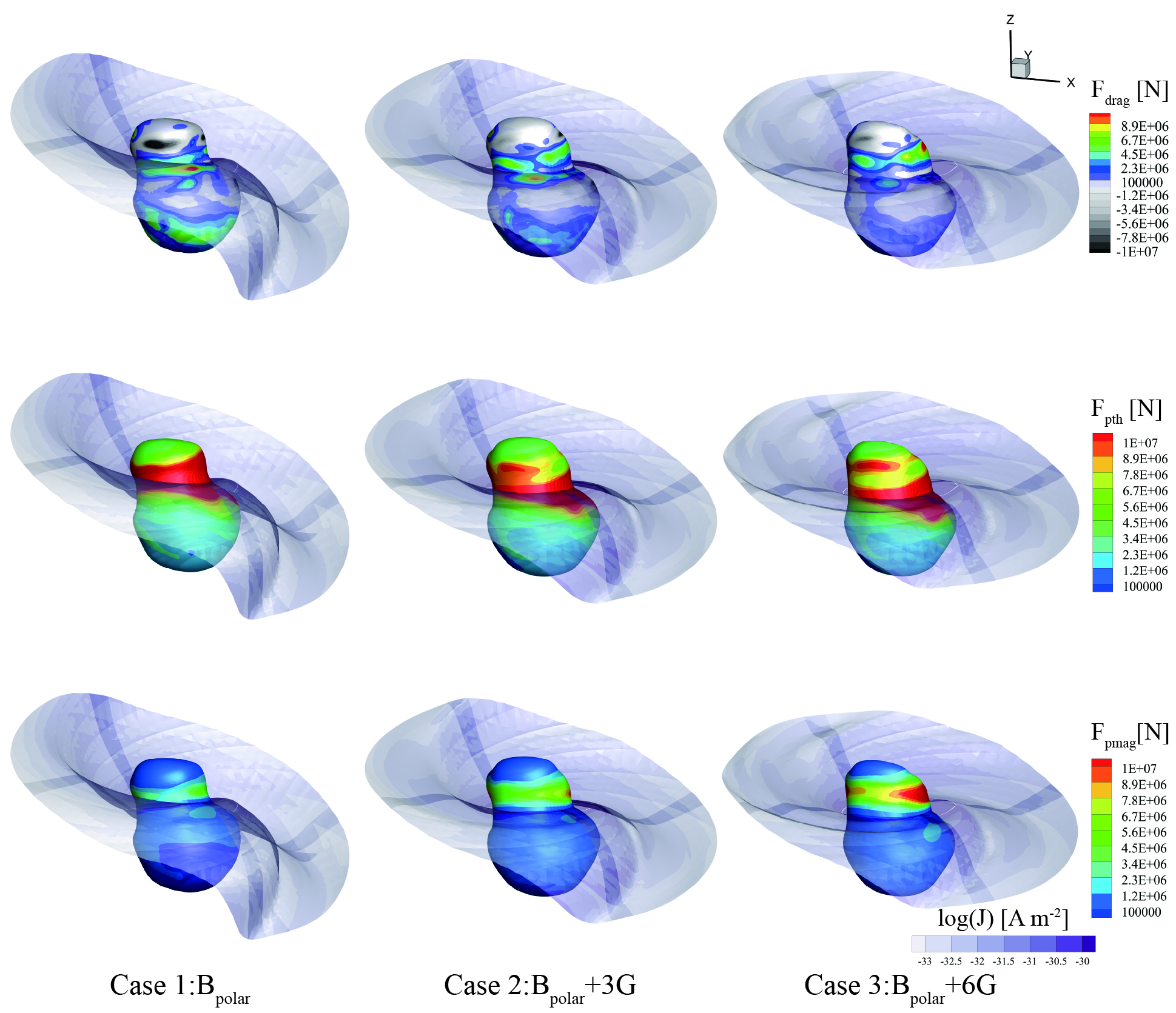}
\caption{Distributions of the external forces per unit area acting on the CME surface under the three magnetic configurations: Case 1 ($B_\textrm{polar}$), Case 2 ($B_\textrm{polar}+3G$), and Case 3 ($B_\textrm{polar}+6G$). The three rows show, from top to bottom, the drag force $F_\textrm{drag}$, the thermal pressure force $F_\textrm{pth}$, and the magnetic pressure force $F_\textrm{pmag}$, each evaluated on the CME boundary. The color indicates the magnitude of each force, while the semi-transparent surface represents the heliospheric current sheet, with the color on the slice indicating the current density $\log(\mid \mathbf{j}\mid)$ (in units of $\mathrm{A \ m^{-2}}$), shown over the same radial range of 20-500 $R_s$.
\label{fig:f10}}
\end{figure*}

Figure \ref{fig:f10} presents the external forces per unit area acting on the CME surface: the drag force (first row), the thermal pressure force (second row), and the magnetic pressure force (third row) under the three magnetic configurations. In this work, the drag force acting on the CME front is computed as $F_{\textrm{drag}}
= \frac{1}{2}\,C_d\,A\,\rho_{\textrm{sw}}\lvert {v}_{\textrm{CME}} - {v}_{\textrm{sw}} \rvert\left( {v}_{\textrm{CME}} - {v}_{\textrm{sw}} \right)$. Here, $C_d$ is the dimensionless drag coefficient, $A$ denotes the contact area between the CME front and the ambient solar wind, $\rho_{\textrm{sw}}$ is the background solar wind mass density. Following previous studies \citep{cargillAerodynamicDragForce2004,vrsnakPropagationInterplanetaryCoronal2013a,sachdevaCMEPROPAGATIONWHERE2015,mayankSWASTiCMEPhysicsbasedModel2024}, $v_{\textrm{CME}}$ and $v_{\textrm{sw}}$ are defined as the full velocity vectors of the CME and the ambient solar wind, respectively, and they are computed locally for every point on the CME surface. For simplicity, we set $C_d = 1$ in this study. The drag force can be understood as the integrated effect of the convective inertial force (or equivalently the dynamic pressure gradient) acting on the CME front.
The pressure-related external forces exerted by the solar wind on the contact area $A$ are given by the thermal pressure force $F_\textrm{pth}=P_{\textrm{th}}\,A$ and the magnetic pressure force $F_\textrm{pmag}=(B_{\textrm{tot}}^2/2)\,A$. 

Figure \ref{fig:f10} shows the distributions of the drag force $F_\textrm{drag}$, the thermal pressure force $F_\textrm{pth}$, and the magnetic pressure force $F_\textrm{pmag}$ per unit area ($A=1$) on the CME surface under the three magnetic configurations.  With increasing polar magnetic field strength ($B_\textrm{polar}$), the magnetic pressure force becomes progressively stronger, while the drag force and the thermal pressure force exhibit only minor variations, indicating that the enhanced background magnetic field provides stronger magnetic confinement of the CME, thereby limiting both its radial and lateral expansion and reducing its propagation.

\begin{figure*}[ht!]
\plotone{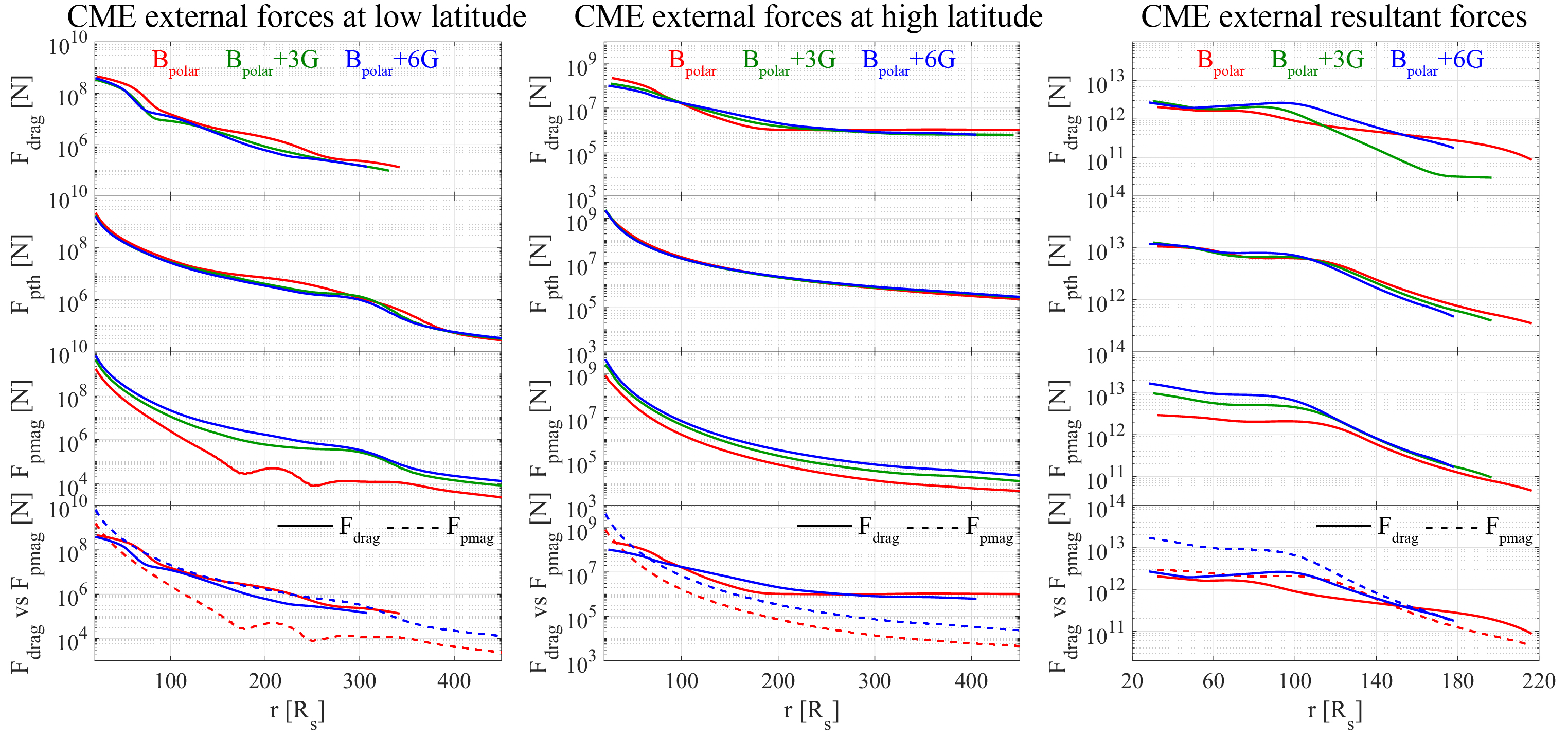}
\caption{Radial profiles of the external forces per unit area acting on the CME surface under the three magnetic configurations: $B_\textrm{polar}$ (red), $B_\textrm{polar}+3G$ (green), and $B_\textrm{polar}+6G$ (blue). The left and middle panels show the external forces extracted at low and high latitudes, respectively, while the right panel displays the resultant forces obtained by summing the external forces per unit area over the CME surface. From top to bottom, each column presents the drag force $F_\textrm{drag}$, the thermal pressure force $F_\textrm{pth}$, the magnetic pressure force $F_\textrm{pmag}$, and the comparison between $F_\textrm{drag}$ (solid lines) and $F_\textrm{pmag}$ (dashed lines).
\label{fig:f11}}
\end{figure*}

Figure \ref{fig:f11} further examines the radial evolution of the external forces per unit area acting on the CME surface for the three magnetic configurations. The results confirm that enhancing the polar field primarily produces large differences in the magnetic pressure force, rather than in the drag or thermal pressure forces.
The comparison between the drag force and the magnetic pressure force (the fourth row) shows that at low latitudes (left panels), the drag force in the $B_\textrm{polar}$ case remains larger than the magnetic pressure force over most of the radial range, whereas in the $B_\textrm{polar}+6G$ configuration the magnetic pressure force becomes comparable to the drag force. At high latitudes (middle panels), the drag force exceeds the magnetic pressure force near 30 $R_s$ in the $B_\textrm{polar}$ case and near 60 $R_s$ in the $B_\textrm{polar}+6G$ configuration, consistent with the conclusion that the magnetic pressure force dominates in the low corona while the drag force becomes increasingly important at larger heliocentric distances \citep{temmerINFLUENCEAMBIENTSOLAR2011,sachdevaCMEDynamicsUsing2017}. 

In addition to the individual force components, we also show the resultant forces (right panels), which are obtained by summing the external forces per unit area over the CME surface. Specifically, the resultant force is calculated by first summing the force components in each Cartesian direction over the CME surface and then taking the magnitude of the resulting force vector, i.e., $
F_{\mathrm{res}}=\sqrt{\left(\sum F_x\right)^2+\left(\sum F_y\right)^2+\left(\sum F_z\right)^2}$
where $F_x$, $F_y$, and $F_z$ denote the local force components per unit area in the $x$, $y$, and $z$ directions, respectively, and the summation is performed over the entire CME surface. For the $B_\textrm{polar}$ case, the resultant drag force surpasses the resultant magnetic pressure force at around 150 $R_s$. This transition is consistent with the results of \cite{zhangStudyingEvolutionICMEs2025}, who reported that CME expansion within roughly 0.7 AU is primarily driven by the magnetic pressure difference between the CME and the solar wind, whereas beyond this distance the expansion is more strongly influenced by the speed difference and associated drag. This behavior is also broadly aligned with the conclusions of \cite{lugazInconsistenciesLocalGlobal2020}. The enhanced polar field makes the radial distance where the resultant drag force overtakes the magnetic pressure force shift outward; in the $B_\textrm{polar}+6G$ case, this crossover is expected to occur at approximately 180 $R_s$. Therefore, under high $B_\textrm{polar}$ conditions, the magnetic pressure force becomes the primary factor responsible for the reduced CME expansion and slower propagation.

\section{Summary and Discussion} \label{sec:highlight}

In this study, we systematically investigate how variations in the photospheric polar magnetic fields influence the Sun-Mars propagation of the 4 December 2021 CME by introducing flux enhancements of the polar magnetic field (increased by 3 G and 6 G) into the background magnetogram. Using the three-dimensional AMR-SIP-CESE MHD model, we simulate the coronal and interplanetary environments corresponding to different polar field strengths and perform a comprehensive analysis of the background solar wind structure, the CME’s propagation speed, its expansion behavior, and the internal and external forces acting on it. Our results reveal the key role of the polar magnetic field in regulating CME evolution throughout the heliosphere.

The simulations show that enhancing the polar magnetic field significantly modifies the background solar wind structure. As the polar field increases from 0 G to 6 G, the high-speed solar wind decelerates by approximately 100 km s$^{-1}$, and the interplanetary magnetic field strength at high latitudes near 1 AU increases by about 0.7 nT. This deceleration is accompanied by an increase in solar wind density, while the thermal pressure remains largely unchanged. These results align with solar-cycle observations linking stronger polar fields to higher-density, stronger-field solar wind. Furthermore, a stronger polar field flattens the heliospheric current sheet and weakens high-speed streams in the ecliptic plane, indicating a more constrained large-scale solar wind topology. Collectively, these changes establish distinct background conditions that directly influence subsequent CME propagation and expansion.

Beyond altering the background solar wind, a stronger polar magnetic field directly suppresses the CME motion. It markedly slows the CME's radial propagation, inhibits its expansion, and alters its arrival time, magnetic structure, and encounter geometry at the spacecraft. For instance, at a given time, the CME in the case with no polar field enhancement has already passed MAVEN/Tianwen‑1 with a broadly distributed magnetic structure, whereas its counterpart under an enhanced polar field remains compact, closer to the Sun, and has not yet reached the spacecraft. Three‑dimensional analysis confirms that stronger polar fields produce a progressively smaller CME with reduced propagation speed, expansion speed, and volume. Quantitatively, an enhancement of the polar magnetic fields with a peak value of 6 G at the pole decreases the mean propagation and expansion speeds by roughly 200 km s$^{-1}$ and halves the CME volume compared to the case without enhancement. Importantly, the slow propagation and reduced expansion of the CME cannot be attributed to the lower background solar wind speed. This is further supported by a controlled experiment in which the background wind speed under the enhanced polar field was restored to the level of the unenhanced case: the suppression of the CME motion persists.

The force analysis further clarifies the physical mechanisms underlying these simulated behaviors. Inside the CME, the thermal pressure gradient consistently dominates over the internal Lorentz force. Within the Lorentz force, the magnetic pressure gradient promotes CME motion, while magnetic tension opposes it. Enhancing the polar magnetic field produces only modest changes to this internal force balance. In contrast, external forces are significantly affected. Strengthening the polar field increases the background magnetic pressure but has little effect on the aerodynamic drag or thermal pressure forces. Consequently, as the polar field strengthens, the relative role of the external drag force gradually diminishes. The region dominated by magnetic pressure shifts outward, and the magnetic pressure gradient becomes the primary factor limiting both the propagation and expansion of the CME.

This study highlights the critical role of the photospheric polar magnetic fields in regulating CME propagation and expansion, and underscores the potential impact of current uncertainties in polar field measurements on space weather forecasting. Although strong polar magnetic fields are expected to slow CME and suppress CME expansion, this study for the first time not only quantifies these effects but also systematically analyzes the underlying physical mechanisms: the magnetic pressure force exerted by the solar wind exceeds the drag force at large heliospheric distances and emerges as the primary factor limiting the CME motion. However, it should be noted that this study focuses on the detailed analysis of a single event. While this case serves as an illustrative example to elucidate the physical mechanisms by which polar fields regulate CME propagation, it is difficult to draw general conclusions about the effect of polar field strength from this isolated event. Future work will build on the current results by applying this methodology to a broader range of CME events to test the generalizability of our findings, and by further examining how the polar magnetic fields influence CME deflection and flux-rope erosion. Given that the current polar magnetic field strength is underestimated \citep{antiochosModelSourcesSlow2011,rileyCanUnobservedConcentration2019}, we artificially enhanced it using a simple method. Future direct observations of the polar region from Solar Orbiter and the Solar Polar Observatory will provide realistic constraints on the polar magnetic fields. These observations will be key to validating our findings and further elucidating the polar field’s influence on the solar wind and CME evolution.

\begin{acknowledgments}
This work is supported by the National Key R$\&$D Program of China (Nos. 2022YFF0503800), China's Space Origins Exploration Program, and the National Natural Science Foundation of China (Nos. 42274213, 42474216, 42530105, 42241118). The work was carried out at National Supercomputer Center in Tianjin, China, and the calculations were performed on TianHe-1 (A).
\end{acknowledgments}


%% For this sample we use BibTeX plus aasjournalv7.bst to generate the

\bibliography{polarcme}{}
\bibliographystyle{aasjournalv7}



\end{document}